\documentclass[aip,apl]{revtex4-1}

\usepackage{graphicx}% Include figure files
\usepackage{dcolumn}% Align table columns on decimal point
\usepackage{bm}% bold math

\begin{document}

\title{Atomic resolution imaging at 2.5~GHz using near-field microwave microscopy}

\author{Jonghee Lee}
\altaffiliation{These authors contributed equally to this work.}
%\email{jonghee@umd.edu}
\affiliation{Department of Materials Science \& Engineering, University of Maryland, College Park, Maryland 20742, USA}

\author{Christian J. Long}
\altaffiliation{These authors contributed equally to this work.}
%\email{crslong@umd.edu}
\affiliation{Department of Physics, University of Maryland, College Park, Maryland 20742, USA}

\author{Haitao Yang}
%\email{HYang@intematix.com}%%
\affiliation{Intematix Corporation, Fremont, California 94538, USA}%

\author{Xiao-Dong Xiang}
%\email{}
\affiliation{Intematix Corporation, Fremont, California 94538, USA}%

\author{Ichiro Takeuchi}
\email{takeuchi@umd.edu}
\affiliation{Department of Materials Science \& Engineering, University of Maryland, College Park, Maryland 20742, USA}

\date{\today} % It is always \today, today,  but any date may be explicitly specified

\begin{abstract}
Atomic resolution imaging is demonstrated using a hybrid scanning tunneling/near-field microwave microscope ($\mu$wave-STM). The microwave channels of the microscope correspond to the resonant frequency and quality factor of a coaxial microwave resonator, which is built in to the STM scan head and coupled to the probe tip. We find that when the tip-sample distance is within the tunneling regime, we obtain atomic resolution images using the microwave channels of the $\mu$wave-STM. We attribute the atomic contrast in the microwave channels to GHz frequency current through the tip-sample tunnel junction. Images of the surfaces of HOPG and Au(111) are presented.
\end{abstract}

%\pacs{68.37.Ef, 68.37.Uv} % PACS, the Physics and Astronomy Classification Scheme.
% 68.37.Ef Scanning tunneling microscopy
% 68.37.Uv Near-field scanning microscopy and spectroscopy

%\keywords{atomic resolution impedance at GHz, near-field microwave microscopy} 

\maketitle
As nanotechnology has rapidly evolved in recent years, scanning probe microscopies are playing an increasingly important role in extracting local physical properties and interactions of materials at the nanometer scale.\cite{rosner:rf_near-field} In this work we report on atomic resolution near-field microwave imaging using a novel scanning probe which combines near-field microwave microscopy (NFMM) and scanning tunneling microscopy (STM). NFMM is most commonly used for mapping the dielectric properties of materials.\cite{Anlage:NFMM_Review,GaoReview2005} In addition, NFMM has potential applications in studying the nonlinearity of tunnel junctions,\cite{tu:021105} spin-sensitive experiments,\cite{manassen:stm-esr,joonhee:microwave_spin-sensitive} and high speed scanning probe imaging.\cite{Kemiktarak2007} NFMM achieves sub-wavelength spatial resolution by bringing the microwave field source to a distance that is much less than the wavelength from the sample under study, overcoming the diffraction-limited spatial resolution barrier.\cite{soohoo:microwave_mag_microscope}

Combining NFMM with STM  provides two advantages. First, STM can be used to bring the probe tip to within a nanometer of a conducting surface. The spatial resolution of NFMM improves as the microwave source gets closer to the material under study. Therefore, the combination of NFMM and STM can potentially enhance the spatial resolution of NFMM. Second, STM can be used to measure electrical properties at sub-angstrom length scales.\cite{binnig:178} This enables simultaneous measurement of microwave near-field interactions and low frequency electrical properties of the sample.

The spatial resolution of the NFMM is traditionally given by the radius of curvature of the sharpened end of the probe tip. The highest demonstrated spatial resolution prior to the present work is 100~nm (Ref.~\onlinecite{gao:1872}). For their work, the tip-sample distance was regulated by a shear force ($< 20$~$\mu$N) between a tip attached to a quartz tuning fork and a sample surface. In our work, we employ a tunnel current ($\sim 1$~nA) to maintain a close contact between the STM tip and a conducting sample surface.

We have looked at two samples for this study: a Au(111) film deposited on mica\cite{sample:Au(111)} and a highly ordered pyrolytic graphite (HOPG) substrate.\cite{sample:hopg-spi} The surface of Au(111) was prepared by flame-annealing at room temperature in air, which removes contamination from the surface and allows surface reconstruction. The surface of HOPG was prepared by cleaving the top layer to expose a clean surface. For both samples we used Pt-Ir STM tips\cite{tip:pt-ir_veeco} because they do not oxidize in ambient conditions. The samples and their preparation methods were selected because of their suitability for STM experiments in ambient conditions.

\begin{figure}[h]
%\centering
%\includegraphics[totalheight=0.3\textwidth]{fig1} % for reprint
\includegraphics[totalheight=0.6\textwidth]{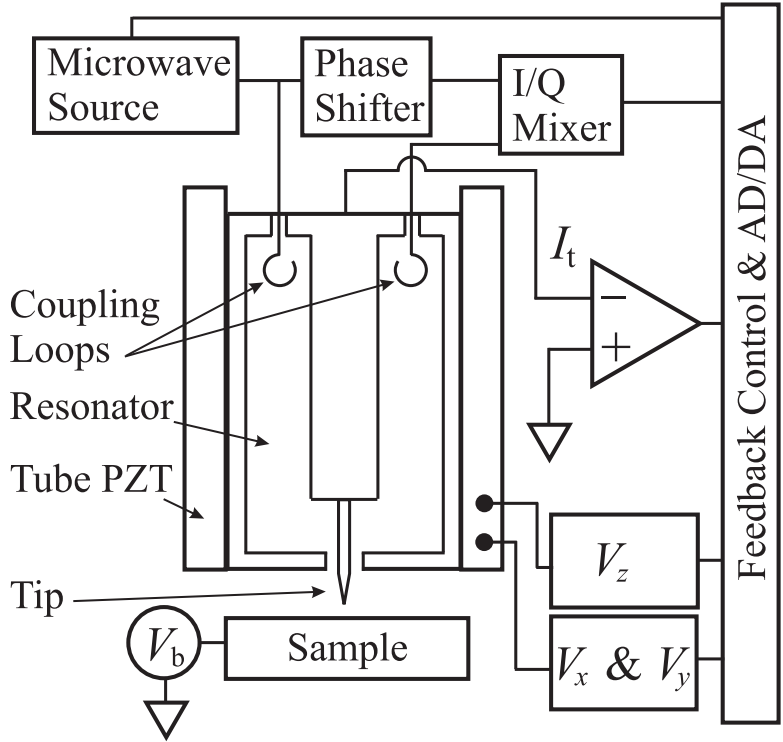} % for manuscript
\caption{\label{fig:esr-stm} Schematic of scanning tunneling/near-field microwave microscope.}
\end{figure}

Fig.~\ref{fig:esr-stm} shows the schematic of our microwave microscope with a built-in STM system. A coaxial $\lambda/4$ resonator is integrated inside the piezo tube scanner. On the open ended side of the resonator, an STM tip protrudes from the resonator to serve as a scanning probe. On the other side of the resonator, there is a directly connected lead for measuring tunnel current as well as input/output ports for microwave signals. Each microwave port consists of a coaxial cable that is inductively coupled to the resonator. The inductive coupling is achieved by forming a single turn loop from the center conductor to the outer conductor of the coaxial lead and placing the loop into the resonator. During a scan, a DC bias voltage is applied to the sample. This experimental set-up allows us to measure the low frequency tunnel current between the tip and sample while simultaneously monitoring the resonant frequency ($f_{\mathrm{r}}$) and quality factor ($Q$) of the resonator.

In order to track the $f_{\mathrm{r}}$ and $Q$ of the resonator during a scan, we employ quadrature homodyne detection.\cite{xiang:microwave_spm_patent} Using this detection scheme allows us to rapidly compute the $f_{\mathrm{r}}$ and $Q$ during a scan without sweeping the microwave driving frequency at each image pixel. The resonator responds to changes in the tip-sample impedance on a time scale given by the decay time of transients in the resonator, $\tau_{\mathrm{decay}}=Q/f_{\mathrm{r}}$. For our system, the upper limit on the bandwidth of the microwave channels is $1 / \tau_{\mathrm{decay}} \approx 4$~MHz, which is far above the bandwidth of our STM (0 -- 10~kHz). Thus, operation of the near-field microscope does not restrict the scan speed of STM.

\begin{figure}[ht]
%\centering
%\includegraphics[totalheight=0.3\textwidth]{fig2} % for reprint
\includegraphics[totalheight=0.6\textwidth]{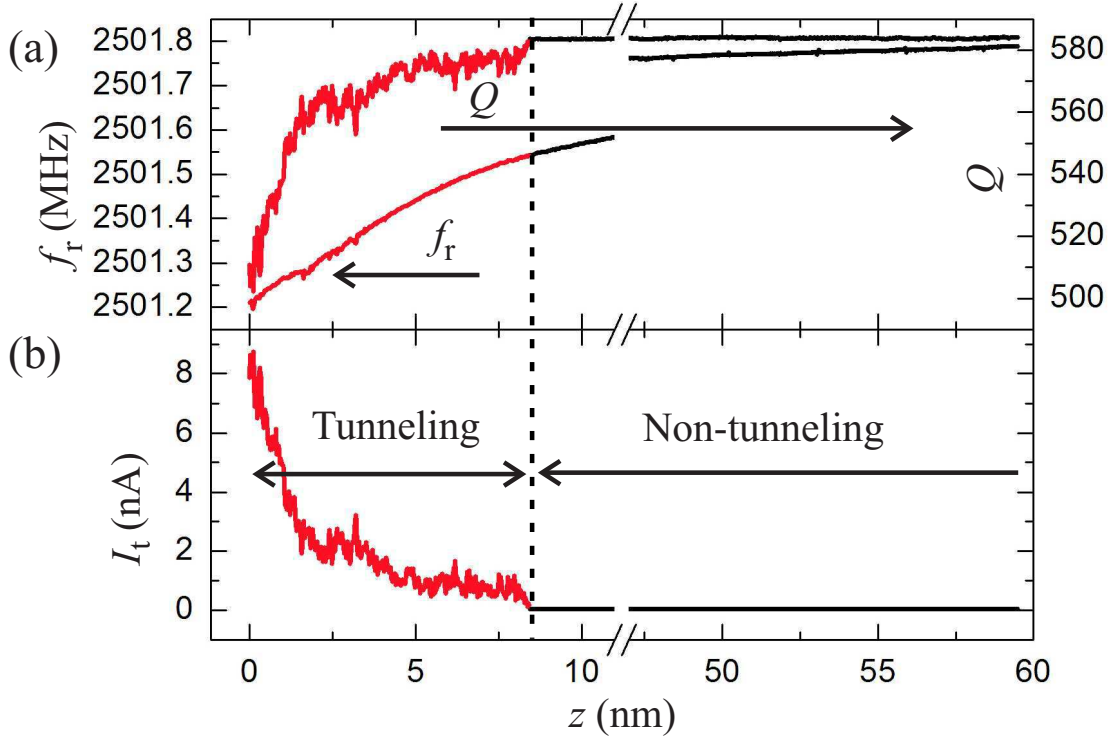} % for manuscript
\caption{\label{fig:approaching_curves} Approach curves of (a) the microwave channels, $f_{\mathrm{r}}$ and $Q$ and (b) STM channel, $I_{\mathrm{t}}$ as as a function of tip height, $z$. Plots are in red when tunneling occurs. All plots are simultaneously acquired on a single tip-sample approach.}
\end{figure}

The unloaded resonator has a nominal $f_{\mathrm{r}}$ of 2.5~GHz and a nominal $Q$ of 600, with some small variation depending on the STM tip that used. As the tip is approached to the sample surface, the microwave near-field interaction between the tip and the sample results in changes in  $f_{\mathrm{r}}$ and $Q$. Figs.~\ref{fig:approaching_curves}a and \ref{fig:approaching_curves}b show the behavior of $f_{\mathrm{r}}$, $Q$, and the tunnel current ($I_{\mathrm{t}}$) as the tip is approached to the sample surface. There are two distinct regimes, one corresponding to the non-tunneling regime (black) and one corresponding to the tunneling regime (red). In the non-tunneling regime, $Q$ is almost unaffected by the tip-sample distance; in the tunneling regime,  $Q$ drops suddenly as $I_{\mathrm{t}}$ increases. The decrease in $Q$ is attributed to dissipation of microwave energy through the tunnel junction. On the other hand, we do not see a slope discontinuity for $f_{\mathrm{r}}$ at the onset of tunneling. This is because $f_{\mathrm{r}}$ depends mostly on the reactive component of the tip-sample junction, which is not affected by the onset of tunneling.

\begin{figure}[ht]
%\centering
%\includegraphics[totalheight=0.4\textwidth]{fig3} % for reprint
\includegraphics[totalheight=0.8\textwidth]{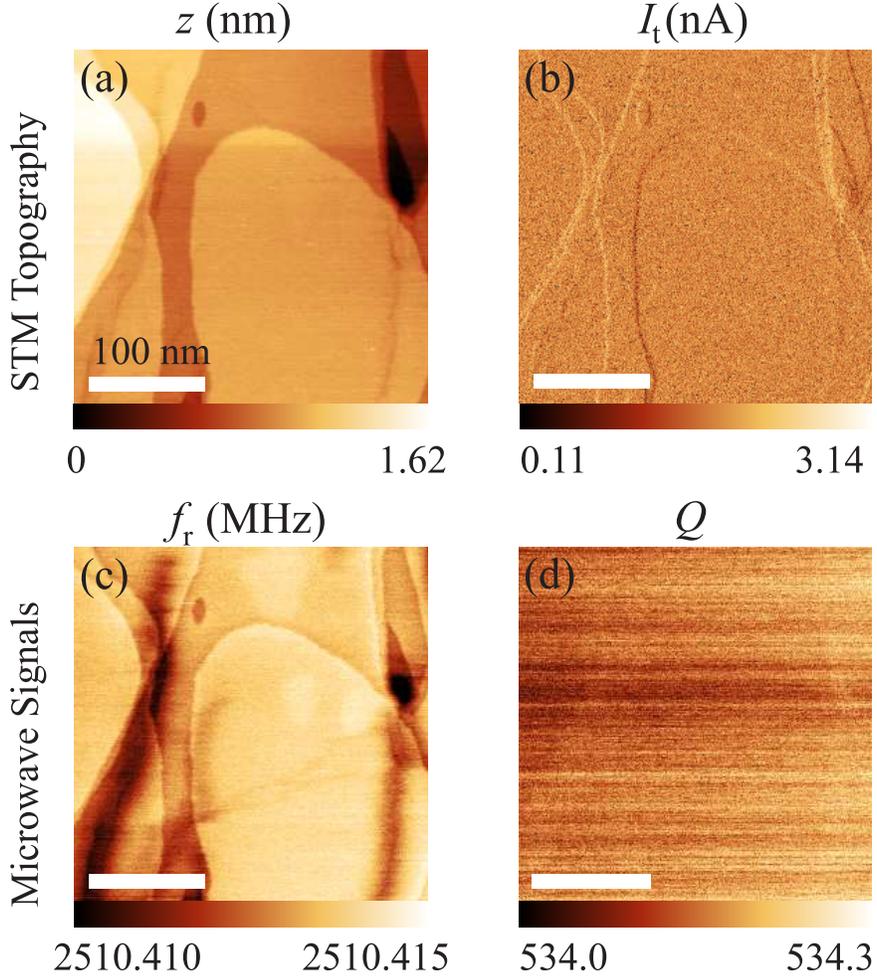} % for manuscript
\caption{\label{fig:gold} Large-scale images of Au(111) taken in STM constant current operation. (a) Tip height, $z$, (b) tunnel current, $I_{\mathrm{t}}$, (c) resonance frequency, $f_{\mathrm{r}}$, and (d) quality factor, $Q$ images.}
\end{figure}

In Fig.~\ref{fig:gold}, we simultaneously collect STM topography and microwave signals while scanning a 300$\times$300~nm$^2$ area of Au(111). Fig.~\ref{fig:gold}a and \ref{fig:gold}b show STM images of tip height ($z$) and $I_{\mathrm{t}}$; Fig.~\ref{fig:gold}c and \ref{fig:gold}d show the microwave signal images of $f_{\mathrm{r}}$ and $Q$ respectively. Data were taken in the STM constant current mode, which regulates the tip-sample distance such that a set point of 0.1~nA (with $V_{\mathrm{bias}}=$100~mV) is maintained. Since the tunnel junction resistance is constant in constant current mode, we observe negligible contrast in the $Q$ channel. On the other hand, the $f_{\mathrm{r}}$ image clearly shows the same surface steps as the STM tip height image. The resonant frequency of the cavity is proportional to $1 \slash \sqrt{L_{\mathrm{eff}}C_{\mathrm{eff}}}$, where $L_{\mathrm{eff}}$ and $C_{\mathrm{eff}}$ are the effective inductance and capacitance of the resonator. Qualitatively, as the tip moves away from the sample, the effective tip-sample capacitance decreases, increasing the resonant frequency of the cavity. Thus, under constant current STM operation, the $f_{\mathrm{r}}$ channel is a convolution of the topography information obtained by STM and the microwave interactions with the sample.\cite{knoll:2667} For highly conducting samples it is possible to take advantage of this fact to perform tip-sample distance feedback using the $f_{\mathrm{r}}$ of a NFMM.\cite{Deuwer1999} However, if one wishes to see variations in materials properties without convolving the sample topography, then it is necessary to scan an atomically flat sample without changing the tip-sample distance.

\begin{figure}[ht]
%\centering
%\includegraphics[totalheight=0.2\textwidth]{fig4} % for reprint
\includegraphics[totalheight=0.4\textwidth]{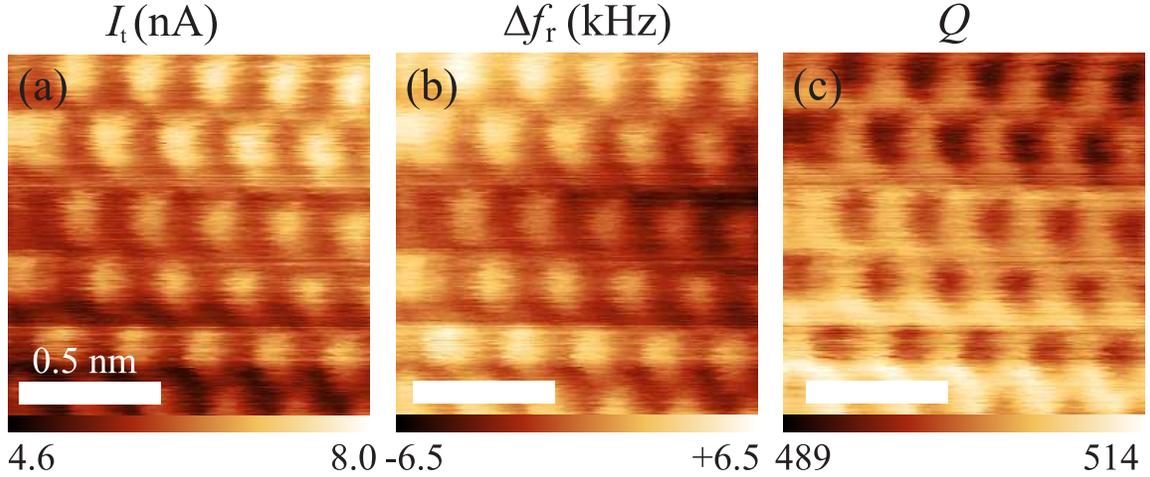} % for manuscript
\caption{\label{fig:atomic_resolution} Atomic resolution images of HOPG taken in STM constant height operation. (a) Tunnel current, $I_{\mathrm{t}}$, (b) resonance frequency shift, $\Delta f_{\mathrm{r}}$, and (c) quality factor, $Q$ images. All images were acquired simultaneously at a scan speed of 20~line/s with a bias voltage of 100~mV. The mean resonance frequency is $\langle f_{\mathrm{r}} \rangle = 2.501486$~GHz.}
\end{figure}

In order to avoid convolution of the NFMM channels with topographic features, we next focused on an atomic-scale area ($\sim$ 1$\times$1~nm$^2$) of HOPG. Scanning a small area allows us to utilize constant height (open loop) mode without damaging the probe tip, while the atomically flat surface of HOPG provides a topography free surface. In order to image HOPG, we first used constant current (closed loop) operation to optimize the bias voltage, tunnel current, and scan speed such that good signal-to-noise was achieved in the STM topography, $f_{\mathrm{r}}$, and $Q$ channels. Once the scan parameters were optimized, we opened the tip-height feedback loop and recorded scan images in constant height mode.

In Fig.~\ref{fig:atomic_resolution}, one can see individual graphite atoms in $I_{\mathrm{t}}$, $f_{\mathrm{r}}$, and $Q$, respectively. For the acquisition of this data set, we used a bias voltage of 100~mV and a scan speed of 20~line/s. The average tunnel current is 6.2~nA with a mean atomic corrugation (peak to trough) of 1.1~nA. The average $f_{\mathrm{r}}$ is 2,501,486.47~kHz with a mean atomic corrugation of 4~kHz. The average $Q$ is 502 with a mean atomic corrugation of 5.5. Using the atomic corrugation of $f_{\mathrm{r}}$, we find the corresponding effective capacitance change of the resonator, $\delta C_{\mathrm{eff}}= 2 C_{\mathrm{eff}} \times \delta f_{\mathrm{r}}/f_{\mathrm{r}} = 8.6 \times 10^{-18}$~F, where $\delta f_{\mathrm{r}}/f_{\mathrm{r}} = 1.6 \times 10^{-6}$ and $C_{\mathrm{eff}} = 2.7 \times 10^{-12}$~F. The $C_{\mathrm{eff}}$ is calculated using the expression in Ref.~\onlinecite{GaoReview2005}. Slight lattice distortion is due to relaxation of the piezo tube scanner during scanning.

In order to verify that we are truly imaging the surface of the HOPG using the microwave impedance between the tip and sample, it is very important to rule out any kind of cross-talk between the STM and microwave channels. In general, there are two possible ways that `artificial' atomic contrast could occur in the microwave channels. The first possible source of cross-talk is through the topography feedback loop, as discussed in regard to Fig.~\ref{fig:gold}. If the tip-sample distance is controlled by the STM tunnel current, and the tunnel current changes as the tip is scanned over some atomic corrugation, then the height of the tip above the sample will also change. As this tip-sample distance varies, the microwave near-field interaction between the tip and sample will also change, resulting in atomic contrast in the microwave channels. This cross-talk mechanism is a particular problem for microwave microscopy because the microwave channels are very sensitive to the tip-sample distance. In order to avoid this scenario, we operated the STM in open-loop mode so that the height of the tip does not depend on the tunnel current.

\begin{figure}[h]
%\centering
%\includegraphics[totalheight=0.2\textwidth]{fig5} % for reprint
\includegraphics[totalheight=0.4\textwidth]{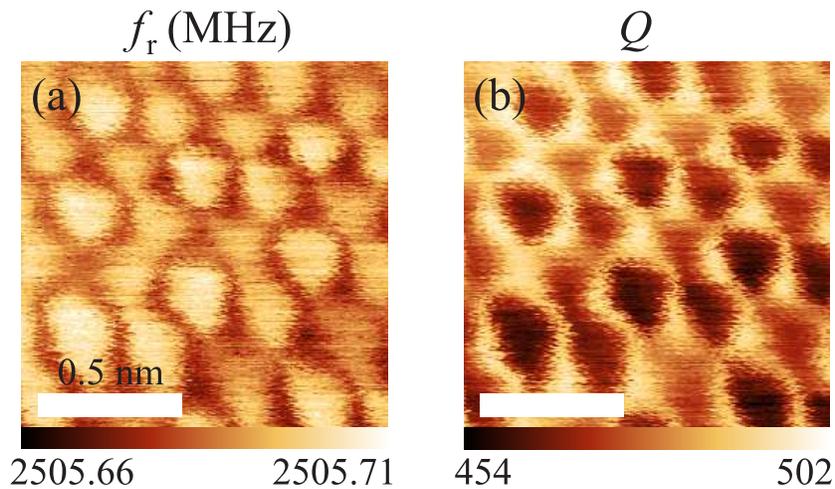} % for manuscript
\caption{\label{fig:gold_atoms} Atomic resolution images of Au(111) taken in only microwave microscopy mode. The STM bias line and the current preamplifier are disconnected from the setup.}
\end{figure}

The second possible source of cross-talk would be some kind of direct coupling between the tunnel current signal and the microwave channels (e.g. capacitive coupling between signal lines). In order to rule out this possibility, we disconnected the DC bias voltage from the sample and disconnected the STM tunnel-current amplifier from the resonator. In this configuration, we approach the tip to the sample using the resonator quality factor as a feedback signal, stopping the tip-sample approach when the quality factor is decreased by about ten percent. Fig.~\ref{fig:gold_atoms} shows atomic resolution images obtained on Au(111) using this method. We found that we were able to obtain atomic resolution images in the microwave channels even without the STM circuitry connected to the system. As a result, the atomic resolution signal observed in the microwave channels can not be caused by cross-talk between the low frequency STM tunnel current and the microwave channels.

By ruling out artifacts caused by STM feedback or signal cross-talk, we conclude that the images of $f_{\mathrm{r}}$ and $Q$ carry information about the impedance of the tunnel junctions at 2.5 GHz. Furthermore, since the atomic contrast occurs only when the tip is within tunneling distance of the sample, we conclude that the atomic contrast in the microwave channels is due to GHz frequency current through the tunnel junction.

In summary, we observe atomic resolution images of HOPG and Au(111) in the microwave channels of our $\mu$wave-STM in ambient conditions, while ruling out artifacts caused by STM feedback or signal cross-talk. Atomic resolution microwave microscopy opens the door to novel ways of probing microwave properties of materials. It is now possible to obtain atomic resolution pictures of materials properties up to the GHz frequency regime, where a variety of interesting element-selective/sensitive phenomena take place including dielectric relaxation and electron spin resonance.\cite{manassen:stm-esr} Moreover, the MHz-bandwidth of the resonator will be useful for high speed spectroscopy measurements at the atomic scale.

\vspace{12pt}
We acknowledge Steven Anlage for useful discussions. This work was supported by the NSF-MRSEC at the University of Maryland, DMR 05-20471 and by the W.~M. Keck Foundation.

%\bibliographystyle{aipnum4-1}
%\bibliographystyle{apsrev4-1}

%\bibliography{atomic_resolution_microwave_signal2}% Produces the bibliography via BibTeX.

%merlin.mbs 2010-03-15 4.21a (PWD, AO, DPC)

\end{document}